# Multilingual research dissemination: Current practices and implications for bibliometrics

**Faizhal Arif Santosa** - National Research and Innovation Agency

**Barbara S. Lancho Barrantes**  - University of Brighton

## Abstract

English is widely used as a lingua franca in scholarly communication, yet preserving local languages is vital to reaching a broader audience. Disseminating research in multiple languages can help ensure equitable access, a responsibility shared by both publishers and authors. This study examines the practices of both groups to identify any notable differences. Several academic social networks, preprint servers, and repositories are analysed to evaluate the resources currently available and their existing policies. Additionally, journals that actively promote multilingual dissemination are reviewed to understand their implementation strategies and how these align with the standards set by the DOI Registration Agency (DOI RA). From the author's perspective, differing policies across platforms can heavily influence decisions, mainly because not all platforms provide relationship metadata. Publishers face similar challenges, underscoring the urgent need for standardisation. Moreover, the lack of consistency creates opportunities for unethical practices in academia, such as counting total of citations originating from the same article in different languages. This highlights the importance of a more comprehensive approach to evaluating research beyond citation and document counts. Collaboration among publishers, authors, and other stakeholders is essential to fostering greater understanding and preventing misconceptions in the academic landscape.

## Introduction

Scientific communication is disseminated on a regular basis, both formally and informally, in a variety of formats, to convey knowledge through a lengthy process, from preprint to post-print as a form of dissemination. This series of activities does not stop when a paper is published; rather, it marks the beginning of reaching a broader audience.[1] Communication will continue to provide space for exchanging ideas and thoughts, where mutual agreement and understanding of the language used are necessary for effective communication. There is a tacit assumption that English will be the language used in academia.[2] Additionally, the view that English is the language of science because most readers understand it[3] and has become the most spoken language in the world[4] positions English as a lingua franca in scholarly communication. This enables people to communicate with colleagues from various countries and languages, fostering collaboration. However, it creates a barrier for people who do not speak English.

---

[1] Garcia et al., "Now Your Manuscript Is Accepted…What's Next?"

Effective communication can ensure the flow of ideas and expertise in both directions.[5] Using English as the sole language of communication can assist in this with language uniformity. However, it is also important to ensure that other languages continue to play a role in scientific communication, considering the language barriers individuals may have. Given that successful dissemination involves reaching an audience with scholarly work that requires the active participation of researchers,[6] communicating to a broad audience without regard for language boundaries is key. Dos Santos et al. highlight the opportunity to enhance the visibility and inclusion of knowledge, while providing assurance to local or regional communities.[7]

Undoubtedly, research funding can come from resources sourced from the public, which creates a sense of responsibility to help the public understand scholarly work by using language that is accessible to funders. The use of the native language tends to be more effective for deep understanding, as well as for the creation and sharing of knowledge.[8] Furthermore, publishing in multiple languages not only broadens the audience but also encourages a diverse perspective on research.[9] Also, the Helsinki Initiative on Multilingualism in Scholarly Communication aims to ensure knowledge is available in various languages to support the dissemination of research results for the benefit of the public.[10] Later, G20 recommendations on inclusion, diversity, and combating inequalities in science, technology, and innovation were announced, which encourage the use of native languages in science.[11]

Providing choices and a variety of languages in scientific communication can facilitate equal access to knowledge for people from all social backgrounds. Considering impact and dissemination, using a language understood by a community will be beneficial.[12] There are good reasons to preserve linguistic diversity in scientific communication amidst the belief that English should be the lingua franca, such as public engagement and generating impact at the local level.[13] Furthermore, publishing research findings in multiple languages can help to develop different perspectives and reach a broader audience.[14] Additionally, with the rapid advancement of technology and the open science movement, academics can reach a wider audience through various alternative media available.[15] For example, the ability to aid "translating" into different languages utilising Large Language Models (LLM), which have outstanding capabilities across varied languages.[16] However, it is essential to note that LLMs require careful user assessment to evaluate answers, especially in translation, considering that the results may be less accurate.[17]

---

[2] Balula and Leão, "Is Multilingualism Seen as Added-Value in Bibliodiversity?"
[3] Dos Santos et al., "The Relationship between the Language of Scientific Publication and Its Impact in the Field of Public and Collective Health."
[4] Ethnologue, "What Are the Top 200 Most Spoken Languages?"
[5] Maryl et al., "The Case for an Inclusive Scholarly Communication Infrastructure for Social Sciences and Humanities."
[6] Garcia et al., "Now Your Manuscript Is Accepted…What's Next?"
[7] Dos Santos et al., "The Relationship between the Language of Scientific Publication and Its Impact in the Field of Public and Collective Health."
[8] Balula and Leão, "Is Multilingualism Seen as Added-Value in Bibliodiversity?"
[9] Kulczycki et al., "Multilingual Publishing in the Social Sciences and Humanities."
[10] Federation of Finnish Learned Societies et al., "Helsinki Initiative on Multilingualism in Scholarly Communication."
[11] Directorate-General for Research and Innovation, "G20 Agree on Open Innovation Strategy and Recommendations for Diversity, Equity, Inclusion, and Accessibility in Science, Technology and Innovation."
[12] Boillos, Bereziartua, and Idoiaga, "The Decision to Publish in a Minority Language."

In this paper, we compare several platforms that provide opportunities for dissemination by exploring what has already been implemented and how specific rules apply to doing so, if any. While the authors use these platforms, we also look at what several publishers have done to provide options for more than one language in their articles to see if there are any differences in practices and how this affects metrics in general. Finally, we evaluate current practices in accordance with the recommendations made by DOI RA.

## Platforms for research dissemination

Sharing translated research results after an article is published is a complex process influenced by multiple factors. The European Code of Conduct for Research Integrity offers guidance on publishing research in alternative forms, such as translations, to clarify their connection to the original work.[18] This practice helps readers understand the original article better while maintaining research integrity and avoiding perceptions of improper dissemination. Licensing plays a crucial role in this process. Articles published under a Creative Commons (CC) licence are easier to translate and share due to their flexible terms. In contrast, works under other licensing agreements present additional challenges, as the ability to share translations depends heavily on each publisher's specific policies. Navigating these considerations is essential to ensure ethical and effective dissemination of translated research.

Currently, there is a wide range of plaftorms options available for research dissemination, each offering distinct advantages, from free platforms to paid services. Academic Social Networks (ASNs), such as Academia.edu and ResearchGate, are widely used for sharing work with a broader audience. ResearchGate offers free services, while Academia.edu provides a more comprehensive paid version. Another popular option is preprint servers, which function similarly to personal web archives for disseminating work.[19] Open Science Framework (OSF) Preprints is an open-source platform that supports many scientific communities, such as SocArXiv, PsyArXiv, and LIS Scholarship Archive. Additionally, arXiv explicitly permits the dissemination of translation works, provided they are valuable to the community.[20] Research can also be shared through various data repositories such as Zenodo, Figshare, and Mendeley Data. For example, Chaleplioglou and Koulouris highlight how Zenodo and OSF Preprints are used as research repositories for ocean and climate science.[21] A study by Sicilia et al. found that in Zenodo, text-based resources, particularly articles, are the dominant type of content shared.[22] These platforms offer diverse methods for disseminating research, with each platform catering to different needs and communities, ensuring that scholars can share their work widely and effectively. Additionally, the DataCite Metadata Properties also provide a mandatory field called "ResourceType" to describe the general type of a resource, with options such as "ConferencePaper" and "JournalArticle".[23] This indicates that certain data repositories support the sharing of more than just research data.

---

[13] Grange, "In All Languages?"
[14] Kulczycki et al., "Multilingual Publishing in the Social Sciences and Humanities."
[15] Garcia et al., "Now Your Manuscript Is Accepted…What's Next?"
[16] Zhao et al., "How Do Large Language Models Handle Multilingualism?"
[17] Zhang et al., "Don't Trust ChatGPT When Your Question Is Not in English."
[18] ALLEA - All European Academies, *The European Code of Conduct for Research Integrity*.
[19] Gao, Wu, and Zhu, "Merging the Citations Received by arXiv-Deposited e-Prints and Their Corresponding Published Journal Articles."

# Methods

Given the wide variety of platforms available, this study focuses on examining dissemination opportunities for translated research from the researchers perspective, specifically on popular non-paid platforms. These include Academic Social Networks (ASNs) like Academia.edu and ResearchGate, preprint servers such as arXiv and OSF Preprints, and data repositories like Figshare, Mendeley Data, and Zenodo. We exclude institutional repositories due to the variability in policies across institutions, although these repositories can also be used. Additionally, we do not consider paid versions of Academia.edu and Figshare+ in this study. To assess the dissemination practices, we reviewed the available guidelines or policies on each platform, where applicable, and uploaded a sample file to explore the fields offered on each platform.

## Academia.edu

Academia.edu is a platform designed for researchers to share their work, connect with peers, and engage with a global academic community. It allows users to upload and disseminate documents, including both academic and non-academic content. For this study, we chose to focus on the free version of Academia.edu, as not all users have access to the paid version of the platform. In general, Academia.edu allows users to freely upload various types of documents, and even non-academic works can be shared. Authors can also utilise multiple languages for their titles, whether by keeping the title in the document's original language, using translated titles, or even combining both. However, there is no specific option to designate a license for the document. Academia.edu also supports multiple file uploads, allowing readers to access additional versions or formats of the same work. It is important to use distinct file names to differentiate between versions of the document (Figure 1). Additionally, the platform supports the use of multiple files with identical digital object identifiers (DOIs) and provides fields for essential publication details, such as the publication name, publisher, year of publication, and abstract, making it a versatile tool for disseminating research.

---

[20] "Translations - arXiv Info."
[21] Chaleplioglou and Koulouris, "Preprint Paper Platforms in the Academic Scholarly Communication Environment."
[22] Sicilia, García-Barriocanal, and Sánchez-Alonso, "Community Curation in Open Dataset Repositories."
[23] DataCite Metadata Working Group, "DataCite Metadata Schema Documentation for the Publication and Citation of Research Data and Other Research Outputs v4.5."

## ResearchGate

ResearchGate (RG) is a social networking platform designed for researchers to share and disseminate their work, collaborate with peers, and engage with the academic community. It serves as a repository for academic content, similar to Academia.edu, but with some distinct features that influence how authors share their research.

On ResearchGate, authors must specify the type of document they are uploading. For example, if the document is a journal article, the "Article" type should be selected; While primarily known as a post print platform, ResearchGate also supports the sharing of preprints and offers the option to generate DOIs for articles that lack one. Additionally, ResearchGate allows authors to use the same DOI as the original version, providing consistency in citation and identification. However, like Academia.edu, ResearchGate does not offer a specific option to assign a license to the uploaded document. Instead, licensing information must be included manually within the document itself.

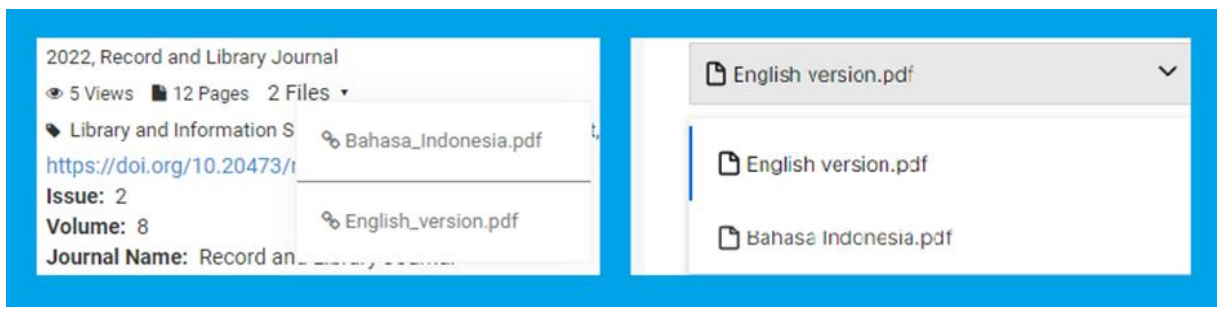

Figure 1. The difference in access to multiple files on Academia.edu (left) compared to RG (right).

## arXiv

arXiv is an open-access preprint server that accepts translations of previously published work if they contribute to the scientific community.[24] Translated titles must follow the format: "A translation of 'TITLE' by AUTHOR," which is less flexible than on platforms like Academia.edu or ResearchGate. Authors are encouraged to use the comments section to indicate the original language or mention if multiple translations exist.[25]

arXiv does not categorise articles by type but automatically generates versioning and supports multiple file uploads. Since 2022, it has offered DOIs linked to the arXiv ID and allows for adding related DOIs for the original manuscript, helping distinguish between versions. arXiv supports six types of Creative Commons (CC) licenses, which must align with the original work's license, especially for ShareAlike licenses (Table 1).

---

[24] "Translations - arXiv Info."
[25] "Non-English Submissions - arXiv Info."

## OSF Preprint

OSF Preprints is a platform for sharing preprints across various academic disciplines, allowing researchers to freely disseminate their work. Unlike arXiv, OSF Preprints offers only CC0 and CC BY licenses for uploaded documents. It does not support uploading multiple files under a single listing; instead, additional files can be added by submitting a new version of the entry.

Publishing information, such as the article type or publication details, is not available on OSF Preprints. There are also no specific title formatting rules, providing flexibility in how titles are presented. However, OSF Preprints shares several features with arXiv, such as straightforward usage and support for metadata and supplementary materials. The submission process consists of six simple steps: filling in the title and abstract, uploading files, entering metadata, completing author assertions, adding supplements (with the option to link to an OSF project), and completing a final review.

## Figshare

Figshare is a platform that allows researchers to share various academic outputs, including articles, data, and supplementary materials. Unlike OSF, which separates preprints from data, Figshare, along with Mendeley Data and Zenodo, combines both types of content in a single repository.

This study focused on the free version of Figshare available at Figshare.com. Users can choose between "Journal Contribution" or "Conference Contribution" as document types when uploading articles. The platform offers CC0 and CC BY licenses for uploaded content. Figshare can generate DOIs for articles, and requesting a DOI identical to the original DOI through the API or by contacting Figshare support is possible.

However, Figshare does not have a dedicated field for publishing information. As a data repository, Figshare supports a relation type field, which allows authors to explain the relationship between two resources by adding various identifier types, such as arXiv, DOI, handle, or URL.

## Mendeley Data

Mendeley Data is a data repository that functions similarly to Figshare, allowing researchers to share datasets, articles, and other types of academic content. Like Figshare, Mendeley Data supports the use of relation type to explain the relationship between resources, though Mendeley currently offers only seven available values for this field.

Mendeley Data generates a reserved DOI for each uploaded item, which can complicate linking to the original source unless the DOI is manually included in another field or within the document itself. Articles uploaded to Mendeley Data can be labeled as either a Research Article or Conference Proceeding, and users can choose from several applicable licenses, including CC0, CC BY, and CC BY-NC. Versioning is automatically handled by Mendeley Data, ensuring that updates and revisions are tracked efficiently.

## Zenodo

Zenodo is a data repository that offers a broader range of licenses for various resource types, particularly for publications like journal articles and conference proceedings. It also provides publication information for related articles, making it more flexible than repositories like Figshare and Mendeley Data.

Zenodo allows depositors to customize version names, such as labeling a file as a "Translated version," which contrasts with the version number format used by the other two repositories (Figure 2). While versioning is typically used for software updates, Zenodo offers more flexibility in this regard. It also includes a field for specifying the language of the content, supporting multiple language selections and eliminating the need for a version statement.

Unlike the DataCite Metadata Schema 4.5, which allows the language field to appear only once,[26] Zenodo enables multiple entries for this field, offering greater customisation. This feature aligns with future updates proposed by DataCite to make this field repeatable and recommended.[27] Zenodo can generate DOIs through its system or use the original DOI, and it supports relationType to refer to the original source. However, as the Metadata Working Group and DataCite continue to refine the IsTranslationOf and HasTranslation properties, these repositories do not yet provide a specific option for describing the relationship between the original and translated versions.

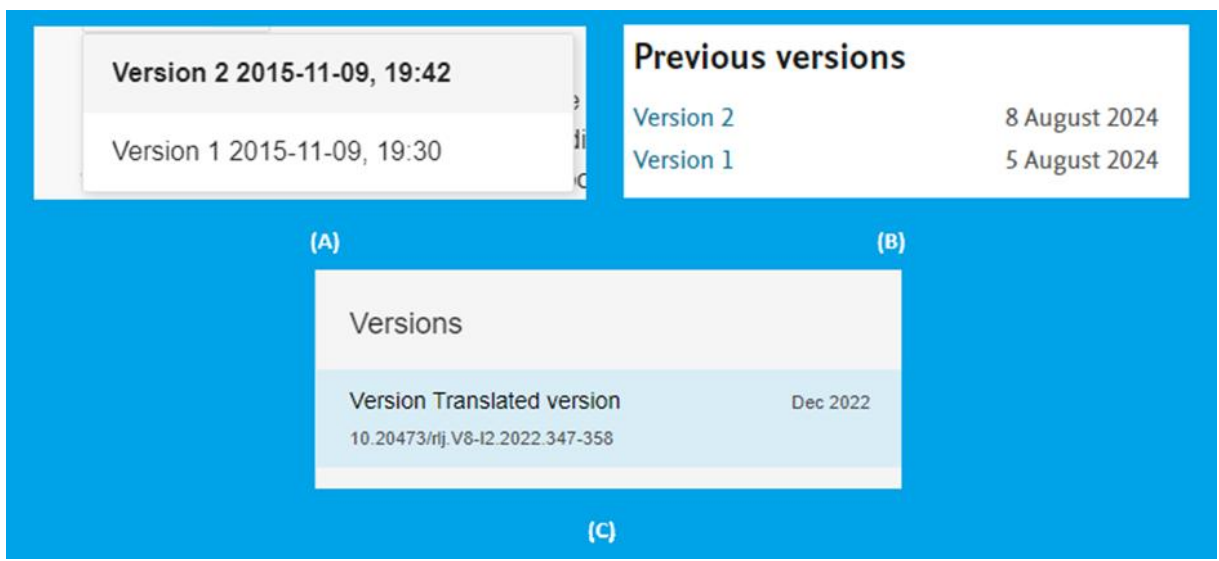

Figure 2. Differences in versioning on Figshare (A), Mendeley Data (B), and Zenodo (C)

---

[26] DataCite Metadata Working Group, "DataCite Metadata Schema Documentation for the Publication and Citation of Research Data and Other Research Outputs v4.5."
[27] Stathis, "Requesting Your Feedback on DataCite Metadata Schema Changes (March 2024)."

Table 1. Comparison of features between each platform

|  | Academia.edu | ResearchGate | arXiv | OSF Preprint | Figshare | Mendeley | Zenodo |
|---|---|---|---|---|---|---|---|
| **Title** | Customisable | Customisable | Should include the title of the translated work and the original author's name | Customisable | Customisable | Customisable | Customisable |
| **Type/ Categories** | Not available | Article or Conference Paper | Not available | Not available | Journal Contribution or Conference Contribution | Research Article or Conference Proceeding | Journal article or Conference proceeding |
| **CC License** | Not available | Not available | CC0, CC BY, CC BY-SA, CC BY-NC-ND, CC BY-NC-SA | CC0, CC BY | CC0, CC BY | CC0, CC BY, CC BY-NC | CC0, CC BY, CC BY-SA, CC BY-NC, CC BY-ND, CC BY-NC-ND, CC BY-NC-SA |
| **Versioning** | Not available | Not available | Automatic | Automatic | Automatic | Automatic | Customisable |
| **Relation Type** | Not available | Not available | Not available | Not available | Support | Support | Support |
| **Identifier** | Support identical DOI to the original | Support identical DOI to the original | Support identical DOI to the original | Support identical DOI to the original | Reserved DOI. Support identical DOI through API or request a ticket to Figshare Support | Not support identical DOI to the original | Support identical DOI to the original |
| **Multiple File** | Allowed | Allowed | Allowed | Not allowed | Allowed | Allowed | Allowed |
| **Publishing Information** | Available | Available | Available | Not available | Not available | Not available | Available |

# Existing Practices

The practice of presenting scientific articles in various languages is not new. However, it has not been widespread due to the many factors that may arise in its implementation, such as the cost of publishing being raised,[28] economic inequality for authors in low-income countries if the responsibility to translate is given to them,[29] and the consistency of language used,[30] as Pisanski explained that it is possible to have different translations between those self-translated by the author and those translated by someone else.[31] We have examined some practices that have been carried out from the publisher's perspective to provide language diversity in a scientific article. Journals that have implemented this multilingual practice have approached this through various models. Some choose to provide language options for all published articles, while others only offer language options for certain articles, indicating a diversity of rules between one and another.

---

[28] Sunol and Saturno, "Challenge to Overcome Language Barriers in Scientific Journals"; Bachelet and Rousseau‑Portalis, "A Technology‑based, Financially Sustainable, Quality Improvement Intervention in a Medical Journal for Bilingualism from Submission to Publication."
[29] Henry Arenas-Castro, "Academic Publishing Requires Linguistically Inclusive Policies."

RDBCI: Digital Journal of Library and Information Science adopted a bilingual format (Portuguese and English) in 2016.[32] Interestingly, RDBCI also includes audio and video content in each published article. Meanwhile, Farmacia Hospitalaria accepts papers in Spanish, has them translated into English, and publishes in both languages.[33] Additionally, BiD, while using Catalan as its working language, still accepts articles in other languages. In fact, in its statement in section 1.2. Languages, it mentions that manuscripts in certain languages will be published in their original language and a Catalan version will be included if possible.[34]

Based on the existing examples, the practice of using diverse languages in publications is not a new phenomenon. However, there are several variations that arise from this practice. RDBCI appears to provide a uniform identifier for all its forms, whether it be the original article, translations, audio, or video, all sharing the same DOI, indicating that despite being in various formats, the entries are still treated as a single item. Meanwhile, BiD, a journal published by the University of Barcelona and Catalonia Open University, has different DOIs for the original article and its translation, although they appear as part of a single record. A more distinct practice is observed in Farmacia Hospitalaria, where the original article and its translation are treated as two separate entries, with the note "Referred to by" on the translation version and "Refers to" to indicate the original article.

From the existing practices, it seems that there is no consensus on whether the DOI should remain the same or should differ. What RDBCI does facilitates information retrieval and also makes it easier for readers to access information in various formats with the same identifier, although there may be potential questions about which source was used. However, with different DOIs – as practised by BiD and Farmacia Hospitalaria – it is easier to indicate which version was cited or used in work. This is useful even though, in practice, individuals generally perform manual translations and still refer to the original article. It is also of use when an article is translated by another publisher; in this case, it is crucial to have different DOIs to identify works from different publishers (even if they are substantively identical), as this will result in different metadata. Another important factor to consider is that both journals must establish a relationship using hasTranslation.[35] For example, the paper by Mavili et al.,[36] published in Turkish by Sosyal Politika Çalışmaları Dergisi, was published in English a few years later by Current Psychology, adding "Translated article:" at the beginning of the title.[37] This practice resembles what arXiv suggests: that translated articles should have a clear statement in the title.

---

[30] Bachelet and Rousseau-Portalis, "A Technology-based, Financially Sustainable, Quality Improvement Intervention in a Medical Journal for Bilingualism from Submission to Publication"; Pisanski Peterlin, "Self-Translation of Academic Discourse."
[31] Pisanski Peterlin, "Self-Translation of Academic Discourse."
[32] "Sobre o Periódico."
[33] "Aims and Scope—Farmacia Hospitalaria | ScienceDirect.Com by Elsevier."
[34] "Submission Guidelines."
[35] Feeney, "Multi-Language Material and Translations."
[36] Mavılı̇, Kesen, and Daşbaş, "Aile Aidiyeti Ölçeği."

## Implications for metrics

Bibliometrics has become a key tool in academia for evaluating the value of published papers, though it typically requires multiple indicators for a comprehensive assessment. Citation indicators have historically been used in the academic system to measure research impact and influence (e.g., publication counts, citation counts, Journal Impact Factor, and h-index).[38] The h-index is a widely used scientometric indicator important for several disciplines that work with a simple combination of the number of publications and citations.[39] This indicator is well known for its simplicity, yet significant differences can occur between disciplines, researchers' age, gender, and indexing services such as Scopus and Web of Science (WoS), etc. This becomes problematic because each citation database calculates metrics based on the data they index, rather than considering all citations. Additionally, Bi highlights four issues with the use of the h-index: it gives equal credit to all co-authors, inflating citation counts, creating imbalances in evaluating research contributions, and opens the door to unethical authorship practices.[40]

This issue also occurs when indexers treat original articles and their translations as separate records, despite having the same content in different languages. For instance, an article in Farmacia Hospitalaria is indexed as two separate entries in Scopus, with one receiving citations while the other does not. The metadata indicates that both have the document type "Article" with the note "[Translated article]" for the translation and identical author keywords and references. Meanwhile, in WoS, both articles have the same title without any additional information to differentiate them. The document type for the translation article is "Editorial Material", and the record for this version does not have any references in its metadata. Even so, other records from this journal show that both the translated and the original have references and the same document type, showing inconsistency.

When an article and its translation are indexed as two separate records, the author ends up with more documents in their profile compared to a publisher who provides only one record for the same article. Citations can affect the author's h-index differently depending on which version of the article is cited. For example, if a citation is made to the original article, it will contribute to the h-index, helping the author reach the required threshold (e.g., an h-index of three). However, if the citation is directed at the translated version, it will only impact the translation's record, not the original article's record. As a result, the author's h-index would remain at two, as illustrated in Figure 3.

---

[37] Mavili, Kesen, and Daşbaş, "Translated Article."
[38] Williams, "What Counts."
[39] Kamrani, Dorsch, and Stock, "Do Researchers Know What the H-Index Is?"
[40] Bi, "Four Problems of the H-Index for Assessing the Research Productivity and Impact of Individual Authors."

Another issue arises when the original article and its translation are treated as separate works. Since the records are considered distinct, each version of the article can receive separate citations, meaning the same work could be counted twice for citations instead of once. This creates opportunities for unethical behaviour, as noted by Chapman, who pointed out that such evaluations can be easily manipulated. Researchers might 'game the system' by artificially inflating citation counts through this duplication, leading to misleading assessments of their work.[41]

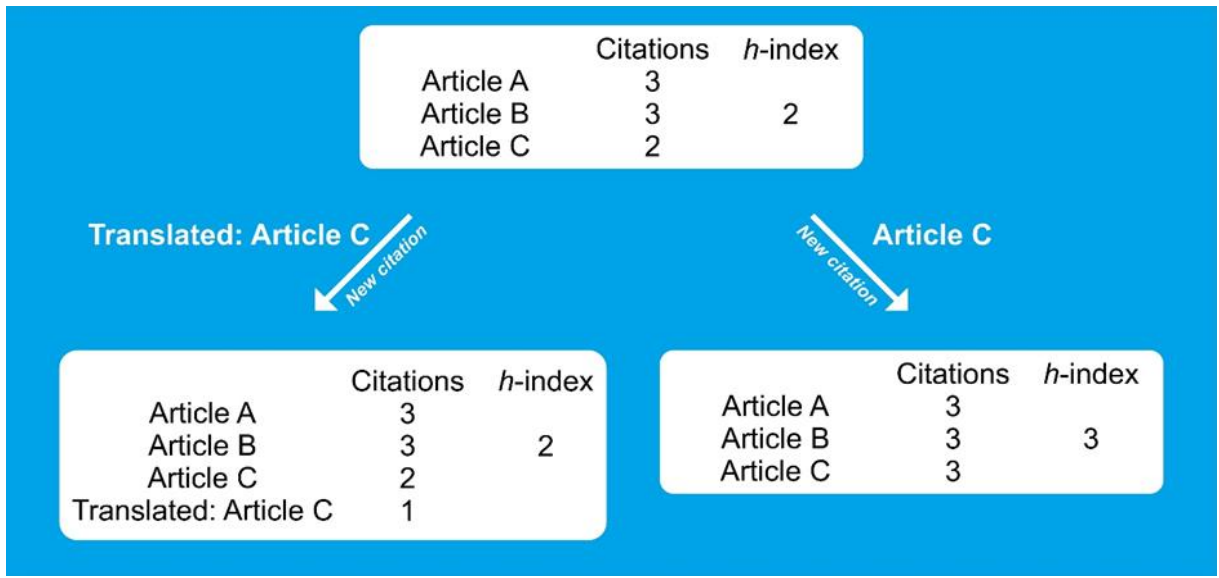

Figure 3. Illustration of potential issues arising with the h-index

Having multiple records for the same work can complicate bibliometric analysis in academia. Since the same keywords are likely to appear in both the original and translated versions of an article, they may be counted multiple times, distorting the analysis. Similarly, when evaluating authors, journals, or institutions, extra effort is required to determine whether the records for the original and translated articles should be treated as one. This requires manually checking each record, as relying solely on duplicate DOIs won't work, since the DOI for the original article and its translation may differ.

To accurately assess performance or other aspects of research, additional steps are required to aggregate citations, ensuring they align with the intended metrics. In citation-based metrics like the Impact Factor (IF), citation inflation is a significant concern. For example, if an article is cited with a different DOI for its translated version, the journal may receive inflated citation counts (Figure 4), especially when compared to sources that publish only in one language or are multilingual. This highlights the need to reconsider how we assess scientific quality, suggesting that peer reviews of the research's social impact could be just as important as citation counts.[42] Furthermore, articles published in national languages often cannot be properly evaluated using traditional article-based metrics,[43] making it necessary to develop more inclusive measurement approaches, particularly for multilingual sources.

---

[41] Chapman et al., "Games Academics Play and Their Consequences."
[42] Wallin, "Bibliometric Methods."
[43] Fuchs, "Bibliometrics: Use and Abuse in the Humanities."

Figure 4. (A) An article received two citations in Dimensions, and (B) a different article also received two citations in Scopus.

# Reflection

Both authors and publishers can provide access to a wider audience through the dissemination of articles in languages other than English. One important consideration for authors is the licensing of their paper. In terms of dissemination, Zenodo is the most flexible option compared to others, but there are certainly other factors to consider when making a choice. Clear attribution to the original article is also necessary, either through a title as suggested by arXiv or by providing a note on the translated version and allowing the use of attribution notes such as the related identifiers available in Mendeley, Zenodo, and Figshare. An overview of multilingual article metadata has been provided by DOI RA such as Crossref and DataCite. Crossref provides guidance for multilingual registration under a single DOI, whereas separate DOIs are assigned when the content is a translation using the hasTranslation relationship.[44] However, Crossref currently states that they still offer limited support for multilingual content,[45] which publishers should consider when transitioning to this model. Meanwhile, based on this explanation, the terms "Refers to" and "Referred to by" are deemed unsuitable, indicating that there is still some misunderstanding about the use of relationships in metadata.

Meanwhile, DataCite, one of the providers of DOIs for research data, has proposed a change in which the language can be repeatable and recommended for use, as well as the emergence of new relation types, HasTranslation and IsTranslationOf,[46] similar to what Crossref offers. For the values of the language, DataCite recommends using IETF BCP 47 and ISO 639-1 language codes. However, in practice on Zenodo, the values "en" and "English" appear, indicating the need for consistency in whether users are shown in code form (en, es, and id) or in full form (English, Spanish, and Indonesian). Authors can carry out this practice, so there is no direct control over the metadata because it is dependent on the authors' understanding, which means that differences and even errors in practice are very likely.

---

[44] Feeney, "Multi-Language Material and Translations."
[45] Farley, "Translated and Multi-Language Materials."

Furthermore, authors must understand that relationship descriptions referencing the original article can be created in the data repository when it becomes available later. However, given each publisher's different policies, it is necessary to confirm whether the publisher is willing to go through additional steps to provide descriptions in other languages that point to other sources. Given the availability of relationType, the description of the relationship on both sides (original version and repository) will help readers understand the relationship created. However, it is possible that the description is only on one side, particularly if the publisher's policy does not address this.

Determining the DOI is also crucial, whether a different DOI is needed or the same as the original article. However, when different publishers or journals are involved, it is essential to have a different DOI, considering that the generated metadata will have an impact. This is in accordance with what Crossref recommends, as it will affect discoverability and citation formatting.[47] Nevertheless, as a means of dissemination, having the same DOI seems important in ASN and data repositories, even though there is an opportunity to generate a new DOI. The lack of agreement or clear guidelines indicates that there are still many paths to take, but caution is needed to avoid falling into the dilemma of self-plagiarism.

# Conclusion

With the goal to reach a wider audience, multilingualism should be taken into consideration by all parties involved in scholarly communication. However, factors such as the quality of writing in different languages, workload, cost, and regulations must be considered before proceeding. Publishers can use the approaches described by Crossref to present it as a single item or, if the content is translated, to explain the relationship using relationship metadata. Meanwhile, if publishers have not yet made space for this practice, authors can use various ASN, preprints, and data repositories with an emphasis on complete metadata, such as relationType for the primary source and field for language description.

However, in practice, there are still differences between what DOI RA recommend for publishers and indexers, such as the use of other relationships and metadata inconsistencies. Another issue that may arise is the various levels of user understanding on this topic, leading to opportunities for inaccuracies. Another factor that could be affected is metrics based on the number of documents and citations, given that there may be differences between the translated scheme and the side that uses multilingualism in its publication. This gap might open up an opportunity for irresponsible actions and raise awareness about the importance of not relying solely on a single metric without considering other factors.

---

[46] Stathis, "Requesting Your Feedback on DataCite Metadata Schema Changes (March 2024)."
[47] Feeney, "Multi-Language Material and Translations."

Publishers and indexers have to collaborate to determine the best formulation involving DOI RA to ensure uniformity of practices in the field, reduce inconsistent actions, and clarify relationships from the readers' perspective. Given that Crossref has previously supported relationship metadata and DataCite's efforts, DOI RA must also work together to support dissemination practices in multiple languages and for various platforms to complete their fields according to the recommendations. Support is available, but attention is required to ensure that various parties can implement this without fear of wrongdoing.

# CRediT author statement

F.A.S.: Conceptualization, Data curation, Investigation, Methodology, Visualization, Writing - original draft, and Writing - review & editing. B.S.L.B.: Conceptualization, Data curation, Investigation, Methodology, and Writing - review & editing.

# Data accessibility statement

Data supporting this study are published in Repositori Ilmiah Nasional https://hdl.handle.net/20.500.12690/RIN/NS4OKB

# Competing interests

The authors have declared no competing interests.